\documentclass[aps,prb,longbibliography,preprint,superscriptaddress]{revtex4-2}
\usepackage{mathrsfs}
\usepackage{amsmath}
\usepackage{amsfonts}
\usepackage{amssymb}
\usepackage{amsthm}
\usepackage{graphicx}
\usepackage{color}
\usepackage{hyperref}
\usepackage{bm}
\usepackage[caption=false]{subfig}
\usepackage{verbatim}

\newcommand{\ak}[1]{\textcolor{black}{#1}}

\begin{document}
\title{Theory of drift-enabled control in nonlocal magnon transport}

\author{Sebastián de-la-Peña}
\address{Condensed Matter Physics Center (IFIMAC), Instituto ``Nicol\'{a}s Cabrera'', and Departamento de F\'{i}sica Te\'{o}rica de la Materia Condensada, Universidad Aut\'{o}noma de Madrid, E-28049 Madrid, Spain} 

\author{Richard Schlitz}
\address{Department of Materials, ETH Zürich, 8093 Zürich, Switzerland}

\author{Sa\"ul V\'elez}
\address{Condensed Matter Physics Center (IFIMAC), Instituto ``Nicol\'{a}s Cabrera'', and Departamento de F\'{i}sica de la Materia Condensada, Universidad Aut\'{o}noma de Madrid, E-28049 Madrid, Spain}

\author{Juan Carlos Cuevas}
\address{Condensed Matter Physics Center (IFIMAC), Instituto ``Nicol\'{a}s Cabrera'', and Departamento de F\'{i}sica Te\'{o}rica de la Materia Condensada, Universidad Aut\'{o}noma de Madrid, E-28049 Madrid, Spain} 

\author{Akashdeep Kamra}
\address{Condensed Matter Physics Center (IFIMAC) and Departamento de F\'{i}sica Te\'{o}rica de la Materia Condensada, Universidad Aut\'{o}noma de Madrid, E-28049 Madrid, Spain} 


\begin{abstract}
Electrically injected and detected nonlocal magnon transport has emerged as a versatile method for transporting spin as well as probing the spin excitations in a magnetic insulator. We examine the role of drift currents in this phenomenon as a method for controlling the magnon propagation length. Formulating a phenomenological description, we identify the essential requirements for existence of magnon drift. Guided by this insight, we examine magnetic field gradient, asymmetric contribution to dispersion, and temperature gradient as three representative mechanisms underlying a finite magnon drift velocity, finding temperature gradient to be particularly effective. 
\end{abstract} 


\maketitle

\section{Introduction}\label{sec:intro}

In ordered magnetic insulators, spin waves can transport spin and heat without an actual motion of electrons~\cite{Akhiezer1968,Kajiwara2010,Kruglyak2010,Bauer2012,Uchida2010,Chumak2015,Nakata2017a}. The corresponding bosonic quasiparticles - magnons - bear several properties distinct from electrons~\cite{Demokritov2006,Takei2014,Sonin2010,Flebus2016,Bozhko2016,Bozhko2019,Brataas2020,Yuan2018,Sonin2020,Kamra2016,Kamra2017}, which make them interesting from both physics and technological perspectives. Due to the generally non-conserved nature of spin, a range of phenomena not admitted by electric current become feasible. Exploiting this principle, coherent spin waves generated using a microwave antenna can be amplified via various mechanisms, such as charge current~\cite{Akhiezer1968,Akhiezer1963,Spector1968}, spin transfer torque~\cite{Seo2009,An2014,Graczyk2021,Padron-Hernandez2011}, and thermal gradient~\cite{Padron-Hernandez2011a,Brechet2013,Yu2017}. Furthermore, attributes such as chirality can be bestowed upon the different spin wave modes via engineering of the energy landscape in the host magnet~\cite{Nembach2015,Gladii2016,Kim2016,An2013,Mook2016,Moon2013,Cortes-Ortuno2013,Garst2017,Tokura2018,Yu2021,Wang2020}.

Spin transport via incoherent magnons, i.e., by a drive such as temperature that does not select a single mode, has also been proposed and demonstrated~\cite{Uchida2010,Cornelissen2015,Bauer2012,Brataas2020}. As a prominent example, nonequilibrium magnons and spin can be injected electrically into a magnetic insulator using a heavy metal electrode~\cite{Zhang2012,Zhang2012a,Cornelissen2015,Goennenwein2015,Li2016,Cornelissen2016,Sinova2015,Velez2016}, made from platinum, for example. Detecting these by a spatially separated electrode provides crucial insights into the nature of spin propagation in the magnet~\cite{Han2020,Takei2014,Zhang2012}. Such a nonlocal method has been inspired, in large part, by the corresponding studies of electronic spin injection and detection in metallic and semiconducting systems~\cite{Johnson1985,Takahashi2003,Fabian2007,Zutic2004}. In the latter, spin chemical potential characterizes spin transport. In a similar fashion, magnon chemical potential~\cite{Cornelissen2016,Bender2012} has been found to play the dominant role in nonlocal spin transport studies employing magnetic insulators. 

Taking further inspiration from this similarity, one may consider the possibility of modulating the magnonic spin propagation length in nonlocal transport studies using pre-existing drift currents. Such effects have previously been demonstrated for electronic spin transport~\cite{Fabian2007,Appelbaum2007,Huang2007,Kameno2014,Ingla-Aynes2016}, and has recently been observed in the context of magnonic spin transport~\cite{Schlitz2021}. Once again, unique properties of these non-conserved magnons should enable novel drift mechanisms associated with, for example, the chirality of eigenmodes~\cite{Nembach2015,An2013,Wang2020,Schlitz2021,Yu2021}.

In this article, inspired by the similarity between nonlocal spin transport in electronic and magnonic systems, we theoretically examine the emergence of drift currents and concomitant control of spin propagation length in magnonic systems. We first establish the drift phenomenology for magnons thereby clarifying its qualitative role in nonlocal magnon transport experiments. Equipped with these insights and using the Boltzmann equation approach~\cite{Kittel2004,Chen2005,Cornelissen2016}, we examine three representative mechanisms that may cause magnon drift - magnetic field gradient~\cite{Meier2003,Kosevich2013,Liao2014,Basso2016,Mook2016,Mook2018,Nakata2015,Nakata2017a}, asymmetric contribution to dispersion~\cite{Gladii2016,An2013,Mook2016,Moon2013,Cortes-Ortuno2013,Schlitz2021,Yu2021,Mook2014,Kim2016,Kim2016a}, and temperature gradient~\cite{Sanders1977,Xiao2010,Adachi2011,Uchida2010,Schreier2013,Bauer2012,Cornelissen2016,Liao2014,Basso2016,Mook2016,Nakata2017a}. These have been studied previously, but without accounting for their drift-mediated coupling to a nonequilibrium magnon chemical potential. We demonstrate that as regards the role of these mechanisms in nonlocal magnon transport, their leading order effect may be captured via a magnon drift velocity. Comparing the latter for these three representative mechanisms, we assess that temperature gradient should offer a good handle for influencing the spin propagation length in nonlocal magnon transport.

\section{Phenomenology of drift}\label{sec:phenomenology}

\begin{figure}[tbh]
	\begin{center}
		\includegraphics[width=85mm]{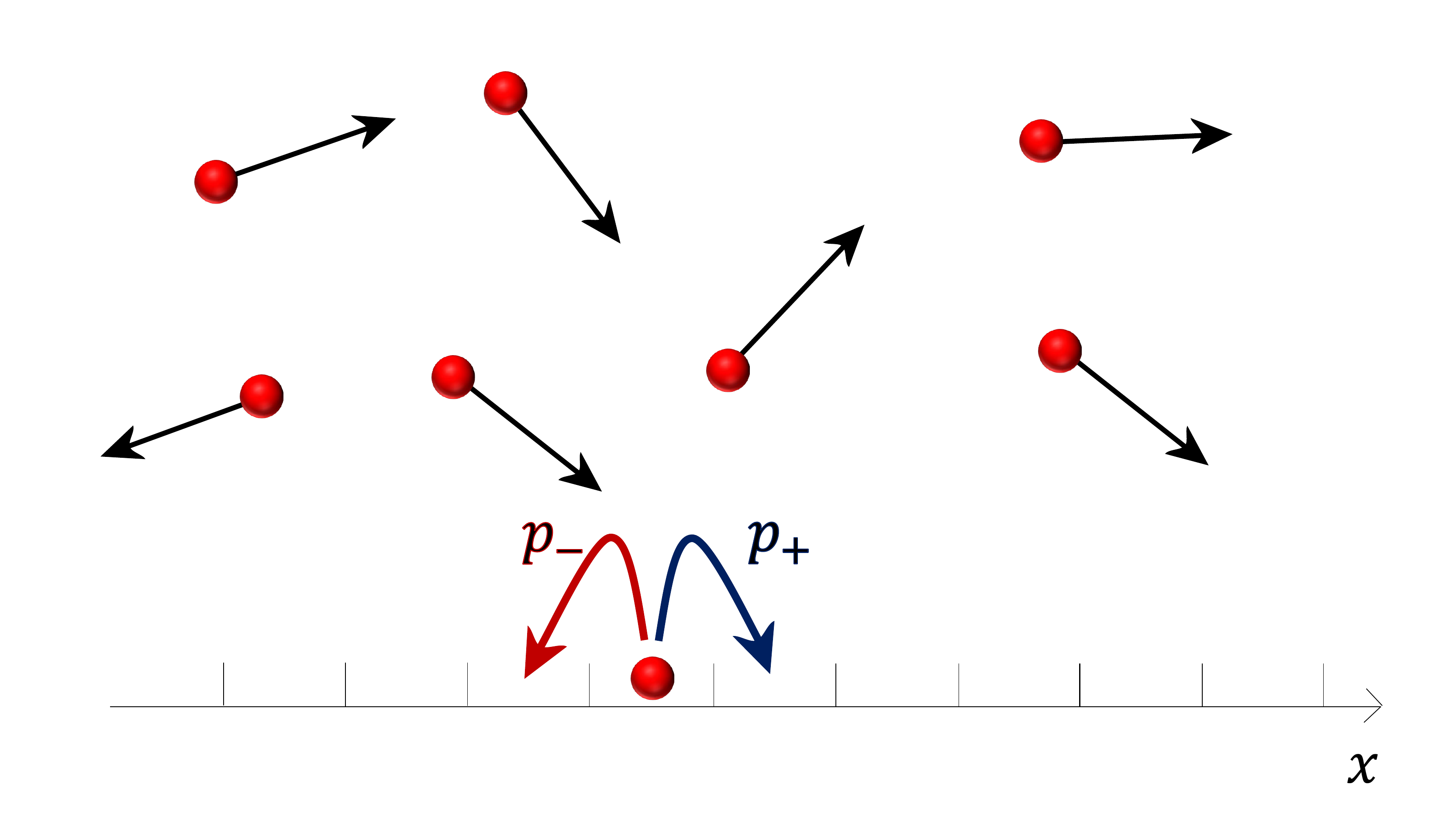}
		\caption{One-dimensional random walk~\cite{Fabian2007}. During one time step of the walk, particles in a bin of size $l$ may move to the right (left) bin with a probability $p_+$ ($p_-$). Their transport bears a drift contribution when $p_+ \neq p_-$.}
		\label{fig:randomwalk}
	\end{center}
\end{figure}

As compared to electronic spin transport~\cite{Fabian2007}, magnons constitute a platform with several qualitative distinctions as spin carriers. These differences include the bosonic, non-conserved nature of magnons as well as their carrying a scalar spin polarized along the direction of equilibrium magnetic order. We begin by considering the one-dimensional random walk~\cite{Fabian2007} of a large number of nonconserved particles to elucidate the requirements for emergence of drift. Dividing the space into bins with a small length $l$, there are $n(x,t)$ particles at a given time $t$ in the bin located at $x$. During one step of the random walk, which is assumed to take a small time $\tau$, the particles can do one of the following three things. They can jump to the right (left) bin with a probability of $p_+$ ($p_-$). Or they can disappear with probability $1 - p_+ - p_-$. Thus, the particles after one step in a bin at position $x$, $n(x,t+\tau)$, were at the adjacent bins before the step such that:
\begin{align}\label{eq:rand1}
n(x,t + \tau) & = p_+ n(x - l,t) + p_- n(x + l,t),
\end{align}
where the first term on the right hand side accounts for the particles that have arrived from the left bin, and the second term accounts for those arrived from the right bin. Employing Taylor expansion in~(\ref{eq:rand1}) and denoting $n(x,t)$ simply by $n$, we obtain
\begin{align} 
n + \frac{\partial n}{\partial t}  \tau& = - \frac{\partial n}{\partial x} l \left( p_+ - p_- \right) + \frac{\partial^2 n}{\partial x^2} \frac{l^2}{2} (p_+ +  p_-) + (p_+ + p_-)n,
\end{align}
\begin{align}
\implies  \frac{\partial n}{\partial t} + \frac{\partial}{\partial x} \left( v_d n - D \frac{\partial n}{\partial x} \right) & = - \frac{n}{\tau_n}, \label{eq:cont}
\end{align} 
where the final result takes the form of particle continuity equation~\cite{Fabian2007}. The current, given by the expression enclosed in brackets [(\ref{eq:cont})], includes a diffusive component parameterized via $D \equiv l^2 (p_+ + p_-)/ (2 \tau)$ and a drift contribution given in terms of the drift velocity $v_d \equiv l (p_+ - p_-) / \tau$. The right hand side of~(\ref{eq:cont}) accounts for particles disappearing with a time constant $\tau_n = \tau / (1 - p_+ -p_-) $. Here, we have assumed the particles to completely vanish at large enough times. In general, relaxation tends to bring the particle density to an equilibrium value~\cite{Kittel2004}. Equation (\ref{eq:cont}) shows that the drift velocity is a direct consequence of particles preferring to go toward one direction as compared to the other. This can be accomplished by breaking the spatial inversion symmetry from the {\em particles' perspective}, which is generally different from the full system Hamiltonian, as clarified further below.

When one considers electrons which are conserved~\cite{Fabian2007}, $p_+ + p_- = 1$ and $\tau_n \to \infty$, such that the sink term on the right hand side of~(\ref{eq:cont}) vanishes. Furthermore, the prototypical drift current in an electronic system results from an applied electric field~\cite{Fabian2007,Appelbaum2007,Huang2007,Kameno2014,Ingla-Aynes2016} such that the drift velocity is proportional to the applied electric field. This relation can be evaluated within the Drude model, for example, leading to Ohm's law. Employing this mechanism, a strong modulation of electronic spin propagation length has been demonstrated in various materials. From a theory perspective~\cite{Fabian2007}, it is often assumed that $p_+ - p_- \ll 1$, which further implies $v_d \ll l/\tau \sim v_{\mathrm{rms}}$, where $v_{\mathrm{rms}}$ is the root mean square velocity of the particles.

In uniformly ordered magnets, a homogeneous applied magnetic field typically does not break the spatial symmetry from the perspective of magnons. Even though it breaks the spatial symmetry for the spin system and consequently the magnetic ground state, the magnon Hamiltonian remains spatially invariant. This has motivated several authors to consider the magnetic field gradient as the driving force for magnon flow~\cite{Meier2003,Kosevich2013,Liao2014,Basso2016,Mook2016,Mook2018,Nakata2015,Nakata2017a}. This mechanism is similar to the case of electric field acting on electrons, which may be seen as the gradient of electric potential~\cite{Chen2005}, and will be addressed in Section~\ref{sec:real}. This is tantamount to breaking the spatial symmetry in real space. Second, we consider the magnon dispersion to bear a contribution antisymmetric in the wave-vector (Section~\ref{sec:reciprocal}). This is representative of symmetry breaking in the reciprocal space, and via the group velocity directionality. This mechanism is intricately related to chiral modes and enables a homogeneous magnetic field to influence the spatial symmetry of magnons~\cite{Gladii2016,An2013,Mook2016,Moon2013,Cortes-Ortuno2013,Schlitz2021,Yu2021}. Finally, we consider the possibility of spatial symmetry breaking by an incoherent property characterizing the system, such as temperature~\cite{Sanders1977,Xiao2010,Adachi2011,Uchida2010,Schreier2013,Bauer2012,Cornelissen2016,Liao2014,Basso2016,Mook2016,Nakata2017a}, in Section~\ref{sec:thermal}. But first, we establish the general role and manifestations of drift in nonlocal magnon transport experiments within a simplified, but rigorous, model in Section~\ref{sec:nonlocal}.

\section{Nonlocal magnon spin transport}\label{sec:nonlocal}

\begin{figure}[tbh]
	\begin{center}
		\includegraphics[width=90mm]{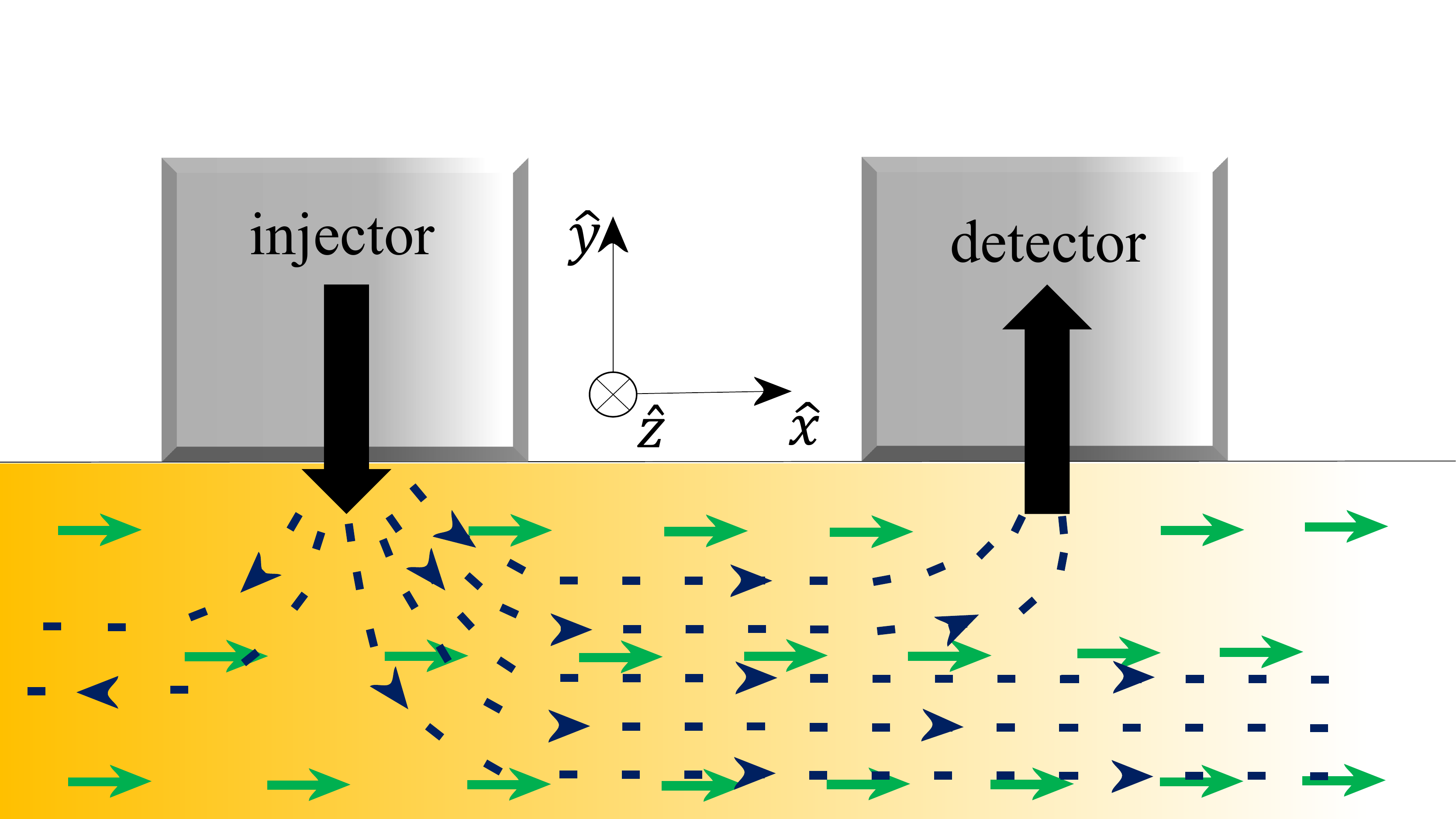}
		\caption{Nonequilibrium magnons, characterized by a nonzero chemical potential, are injected into a ferromagnetic insulator (yellow) by electronic spin accumulation generated in a heavy metal electrode (gray). These transport through the magnet and are detected by a separate heavy metal electrode as the magnon chemical potential at its location. Due to a pre-existing drift current (depicted via green arrows) in the magnet, the magnon propagation length bears a contribution proportional to the drift current [(\ref{eq:lp})] allowing for the former's control via the latter.}
		\label{fig:setup}
	\end{center}
\end{figure}

We now describe our general model for magnonic spin transport observed via electrical injection and detection of spin in a ferromagnetic insulator using the setup depicted in Figure~\ref{fig:setup}. Via the spin Hall effect~\cite{Hirsch1999,Sinova2015}, a charge current driven through the injector electrode generates spin accumulation at its interface with the magnet, which in turn injects magnons into the latter. These propagate in the magnet reaching detector electrode on the right, where they drive a spin current into the detector. Via inverse spin Hall effect~\cite{Saitoh2006,Sinova2015}, this is detected as a voltage under open circuit conditions. Thus, a charge current in the injector leads to a voltage in the detector mediated by magnonic spin transport in the magnet. The ratio of this detected voltage to the injected current is termed nonlocal magnetoresistance and is directly measured in experiments~\cite{Zhang2012,Cornelissen2015,Goennenwein2015,Cornelissen2016,Li2016}. 

Focusing on the magnon propagation in the ferromagnetic insulator first, the transport can be formulated in terms of the magnon continuity, derived using the Boltzmann equation approach~\cite{Kittel2004,Chen2005,Cornelissen2016} in \ref{sec:formulate}:
\begin{align}
\frac{\partial n}{\partial t} + \pmb{\nabla}\cdot\pmb{j}_m & = - \frac{n - n_0}{\tau_m}, \label{eq:magcont}
\end{align}
where $\pmb{j}_m$ is the magnon current density, $\tau_m$ is the magnon decay time, and $n_0$ is the equilibrium magnon density. As detailed further in \ref{sec:formulate}, the relatively simple formulation of~(\ref{eq:magcont}) is enabled by the separation of magnon scattering time scales~\cite{Cornelissen2016}. The processes that conserve the spin and consequently the number of magnons are predominantly mediated by exchange interaction, which is the strongest energy scale in the magnet, and are fast~\cite{Akhiezer1968,Cornelissen2016}. We characterize them with a small phenomenological time $\tau$. The value of $\tau$ and the scattering processes that come to dominate it will depend on the situation under consideration. On the other hand, the magnon decay takes places via processes that do not conserve their number and that are mediated by weaker interactions, such as dipole-dipole interaction~\cite{Akhiezer1968,Cornelissen2016}. These are characterized by a larger time scale $\tau_m \gg \tau$. The situation is thus similar to the case of random walk considered in the preceding section.

Due to this hierarchy of time scales, primarily the nonzero magnon chemical potential characterizes the excess magnons (injected by a heavy metal, for example) and their transport in the magnet~\cite{Cornelissen2016}. Furthermore, due to a rapid thermalization without changing the number of magnons, the latter can be characterized by a common local temperature for the entire system that includes phonons. As a result of this, the magnon density time dependence \ak{can primarily be described via the chemical potential, as detailed further in \ref{sec:formulate}}:
\begin{align}\label{eq:mudot}
\frac{\partial n}{\partial t} & = \chi \frac{\partial \mu}{\partial t},
\end{align} 
where the general form of $n(\pmb{r},t) = n_{0} + \chi \mu(\pmb{r},t)$ \ak{is obtained using the Boltzmann theory formulated in \ref{sec:formulate}} for the various specific cases considered in this work. Further, the magnon current density can be expressed in the following general form:
\begin{align}\label{eq:jgen}
\pmb{j}_m & = \pmb{j}_0 - \sigma_m \pmb{\nabla} \mu + \chi \pmb{v}_d \mu,
\end{align}
where $\pmb{j}_0$ is a spatially homogeneous contribution independent of $\mu$, $\sigma_m$ is the magnon conductivity~\cite{Cornelissen2016}, and $\pmb{v}_d$ becomes the drift velocity. \ak{The specific expressions for the magnon current density $\pmb{j}_m$ induced by different drives are obtained in different sections below using the formalism described in \ref{sec:formulate}.} Combining Eqs.~(\ref{eq:magcont})-(\ref{eq:jgen}), we obtain the equation governing magnon chemical potential and thus, transport:
\begin{align}\label{eq:chemcont}
\frac{\partial \mu}{\partial t} - D_m \nabla^2 \mu + \pmb{v}_d \cdot \pmb{\nabla} \mu & = - \frac{\mu}{\tau_m},
\end{align}
where $D_m = \sigma_m/\chi$ is the magnon diffusion constant. We treat~(\ref{eq:chemcont}) as the definition for drift velocity as the equation allows its direct interpretation in terms of its effect on magnon transport, as demonstrated below. In the following sections, we evaluate and compare $\pmb{v}_d$ under various situations hosting finite drift currents.

In order to capture the qualitative physics within a simplified analytic model~\cite{Kamra2020,Cornelissen2016}, we assume the magnetic layer to be thin and the device depicted in Figure~\ref{fig:setup} to be spatially invariant along the $z$-direction. With these assumptions, we may assume the magnon chemical potential to depend only on the $x$ coordinate making the equation to be solved [~(\ref{eq:chemcont})] one-dimensional. \ak{This approximation is tantamount to disregarding chemical potential and temperature variation in the direction perpendicular to the magnet/heavy metal interface. While being a good approximation for thin films, it breaks down when the magnetic layer thickness is comparable to the magnon diffusion length of few microns. Under those conditions, a two- or three-dimensional numerical modeling is necessary~\cite{Cornelissen2016}.} We further assume the injector and detector electrodes to be weakly coupled with the magnetic insulator layer, and the magnon drift velocity $\pmb{v}_d$ to be directed in the magnetic film plane. Under these circumstances, as detailed in \ref{sec:Rnl}, the magnon injection is accounted for via the boundary condition~\cite{Kamra2020}:
\begin{align}\label{eq:bc1}
- \left. \sigma_m \frac{\partial \mu}{\partial x} \right|_{x=0} & = j_{\mathrm{in}},
\end{align}  
where we assume the injector to be located at $x = 0$, and $j_{\mathrm{in}}$ is proportional to the charge current driven through the injector electrode as detailed in \ref{sec:Rnl}. With these assumptions and in steady state, the solution to~(\ref{eq:chemcont}) for the range $x>0$ is obtained as:
\begin{align}
\mu(x) & = \frac{l_p j_{\mathrm{in}}}{\sigma_m} \exp \left( - \frac{x}{l_{p}} \right), \label{eq:muevol} \\
l_{p} & = \sqrt{\lambda_m^2 + \left( \frac{v_{dx} \tau_m}{2} \right)^2} + \frac{v_{dx} \tau_m}{2} \approx \lambda_m +  \frac{v_{dx} \tau_m}{2}, \label{eq:lp}
\end{align}
where $\lambda_m \equiv \sqrt{D_m \tau_m}$ is the magnon diffusion length, $v_{dx}$ is the $x$ component of the drift velocity, we assume $|v_{dx} \tau_m| \ll \lambda_m$ for simplicity, and we further required $\mu(x \to + \infty) \to 0$ in obtaining the above solution. Equation (\ref{eq:lp}) shows that the magnon propagation length $l_{p}$ bears a contribution from drift, which can be positive or negative depending on the sign of $v_{dx}$. Thus, the magnon propagation can be modulated via the drift~\cite{Schlitz2021}. Without loss of generality, we consider the drift to point along the $x$ axis in the rest of our analysis such that we may replace $v_{dx} \to v_d$. 
 
Within our assumption of a weakly coupled detector electrode, as detailed in \ref{sec:Rnl}, the nonlocal voltage signal detected by the right heavy metal electrode is proportional to the magnon chemical potential at its location $x_d$~\cite{Kamra2020,Cornelissen2016}:
\begin{align}
\mu(x_d) & = \frac{l_p j_{\mathrm{in}}}{\sigma_m} \exp \left( - \frac{x_d}{l_{p}} \right), \\
   & \approx \frac{\lambda_m j_{\mathrm{in}}}{\sigma_m} \exp \left( - \frac{x_d}{\lambda_m} \right) \left( 1 + \frac{v_d \tau_m}{2 \lambda_m} + \frac{x_d v_d \tau_m}{2 \lambda_m^2} \right), \label{eq:muxd}
\end{align}
where we assumed $|x_d v_d \tau_m| \ll \lambda_m^2$ in addition to $|v_d \tau_m| \ll \lambda_m$ assumed already. Equation ~(\ref{eq:muxd}) shows that a positive drift velocity increases the signal detected by the right heavy metal electrode. The contribution of drift can be isolated experimentally by recording the dependence of this signal [(\ref{eq:muxd})] on handles, such as applied magnetic field direction~\cite{Schlitz2021} or temperature gradient magnitude, that influence the drift velocity.

In typical experiments, the heavy metal electrodes are coupled not so weakly to the magnetic insulator since the detected signal is directly proportional to the square of this coupling, quantified via the interfacial spin conductance~\cite{Cornelissen2016}. Under those conditions, the analysis of experimentally measured nonlocal magnetoresistance becomes tedious and simple analytic expressions are not available~\cite{Cornelissen2016}. Hence, while our analysis captures the qualitative physics well, it might become necessary to perform simulations for a comparison with experiments. Further, our thin-film one-dimensional model does not admit the possibility of different drifts at the two surfaces of the magnet. In case the drift current varies through the magnetic insulator thickness, the magnon chemical potential transport is more sensitive to the drift current close to the surface that is interfaced with the heavy metal electrodes. Such effects can be captured by solving the full 3-dimensional~(\ref{eq:chemcont}) and are beyond the scope of this work.

\section{Real space dependence of dispersion}\label{sec:real}

We now derive expressions for magnon current and the corresponding drift velocity considering a spatially dependent dispersion. An example of this situation is an applied magnetic field with its magnitude varying linearly in space. Such a drive has been considered by several authors~\cite{Meier2003,Kosevich2013,Liao2014,Basso2016,Mook2016,Mook2018,Nakata2015,Nakata2017a} previously using a broad range of analytic and numerical methods. Here, we follow the Boltzmann equation approach of Ref.~\cite{Liao2014}, in which the focus was on thermal transport of the magnons driven by the magnetic field gradient. Generalizing their methodology~\cite{Liao2014,Chen2005,Kittel2004} to include a nonzero magnon chemical potential~\cite{Cornelissen2016}, we obtain the desired expression for the magnon current and drift velocity.

In this approach, a WKB approximation is made and magnons are assumed to have an energy $\hbar \omega$ that depends on the spatial position: $\omega \equiv \omega(\pmb{r},\pmb{k}) = \omega_H(\pmb{r}) + \omega_0 + \beta k^2 \equiv \omega_H(\pmb{r}) + \omega_s(\pmb{k})$. Here, we consider the spatially dependent part to result from the Zeeman energy due to the applied magnetic field $H(\pmb{r}) \hat{\pmb{x}}$: $\omega_H(\pmb{r}) = |\gamma| \mu_0 H(\pmb{r})$, where $\gamma$ is the typically negative gyromagnetic ratio, and $\mu_0$ is the permeability of vacuum. The contribution $\omega_s = \omega_0 + \beta k^2$ results from anisotropy and exchange interaction. As outlined in \ref{sec:formulate}, the dependence of magnon current on the drives is obtained by solving the Boltzmann equation at short time scales corresponding to the time $\tau$. The magnon current becomes (\ref{sec:formulate}):
\begin{align}\label{eq:jm}
\pmb{j}_m & = \int_{\mathrm{BZ}} \frac{d^3k}{\left( 2 \pi \right)^3} \pmb{v}_g f(\pmb{r},\pmb{k}) , 
\end{align}
where $\pmb{v}_g = \pmb{\nabla}_{\pmb{k}} \omega = 2 \beta \pmb{k}$ is the group velocity, and $f(\pmb{r},\pmb{k})$ is the distribution function given in the relaxation time approximation by (\ref{sec:formulate})
\begin{align}\label{eq:fmaggrad}
f(\pmb{r},\pmb{k}) & = n_{B}(\hbar \omega - \mu) + \delta f(\pmb{r},\pmb{k}), \quad \mathrm{with} \\
 \delta f(\pmb{r},\pmb{k}) & =  - \tau \left. \frac{\partial n_{B}}{\partial \epsilon} \right|_{\hbar \omega - \mu} \pmb{v}_g \cdot \left( \hbar \pmb{\nabla} \omega_{H} - \pmb{\nabla} \mu  \right),
\end{align}  
where $n_B(\epsilon) = 1/(\exp(\epsilon/(k_B T)) - 1)$ is the Bose distribution function with $k_B$ the Boltzmann constant and $T$ the temperature. Retaining only linear contributions, we obtain from Eqs.~(\ref{eq:jm}) and (\ref{eq:fmaggrad}):
\begin{align}\label{eq:jmmaggrad}
\pmb{j}_m & = \chi D_m |\gamma| \hbar \mu_0 \pmb{\nabla} H - \chi D_m \pmb{\nabla} \mu + \chi 2 \beta \tau |\gamma| \mu_0 \Gamma ~ \mu ~ \pmb{\nabla} H,
\end{align}
where 
\begin{align}
\chi & \equiv  - \int_{\mathrm{BZ}} \frac{d^3k}{\left( 2 \pi \right)^3}  \left. \frac{\partial n_{B}}{\partial \epsilon} \right|_{\hbar \omega_s}, \label{eq:chi} \\
 D_m & \equiv - \frac{4 \beta^2 \tau}{\chi}  \int_{\mathrm{BZ}} \frac{d^3k}{\left( 2 \pi \right)^3}  \left. \frac{\partial n_{B}}{\partial \epsilon} \right|_{\hbar \omega_s} k_z^2 , \label{eq:Dm} \\
 \Gamma & \equiv \frac{2 \beta \hbar}{\chi}  \int_{\mathrm{BZ}} \frac{d^3k}{\left( 2 \pi \right)^3}  \left. \frac{\partial^2 n_{B}}{\partial \epsilon^2} \right|_{\hbar \omega_s} k_z^2 . \label{eq:Gamma}
\end{align}
A comparison of the magnon current expression thus obtained [(\ref{eq:jmmaggrad})] with its general form~(\ref{eq:jgen}) enables identification of the different contributions. Further, employing~(\ref{eq:jmmaggrad} in the magnon continuity equation (\ref{eq:magcont}) leads us to~(\ref{eq:chemcont}) with
\begin{align}\label{eq:vdmaggrad}
\pmb{v}_d & = 4 \beta \tau \Gamma |\gamma| \mu_0 \pmb{\nabla} H.
\end{align}
In obtaining our result in the form of~(\ref{eq:chemcont}), a small correction to $D_m$ has been disregarded. Furthermore, the drift velocity given by~(\ref{eq:vdmaggrad}) that enters the magnon continuity equation (\ref{eq:chemcont}) is larger by a factor of 2 compared to the drift velocity that enters the current density expression [(\ref{eq:jmmaggrad})]. This additional contribution results due to the magnetic field dependence of $D_m$ and the consequent term in the divergence of magnon current density.

Equation (\ref{eq:vdmaggrad}) is the main result of this section. \ak{We note that the smaller magnon scattering time $\tau$ determines the drift velocity [equation (\ref{eq:vdmaggrad})], instead of the magnon decay time $\tau_m$.} Among the various factors that influence drift velocity, only $\tau$ bears a substantial temperature dependence making low temperatures favorable for higher drift velocities. In order to estimate the latter at room temperature, we consider $\Gamma \approx 1$, $\tau = 1$ ps~\cite{Cornelissen2016}, $\hbar |\gamma|/ k_B = 1.3$ K/T~\cite{Liao2014}, $\hbar \beta = 8 \times 10^{-40}$ J$\mathrm{m}^2$~\cite{Cornelissen2016} and obtain $\pmb{v}_d \approx  ( 5 \times 10^{-6} ~\mathrm{m}^2/\mathrm{s} \mathrm{T} ) ~ \mu_0 \pmb{\nabla} H$. Here, magnon-phonon scattering is expected to dominate $\tau$ which is approximated as the inverse magnon-phonon scattering rate~\cite{Cornelissen2016}. Thus, rather large magnetic field gradients are required for even modest drift currents. 

\section{Reciprocal space dependence of dispersion}\label{sec:reciprocal}

We now consider a mechanism for drift that emerges from spatial symmetry breaking in the reciprocal space. This ensues when the dispersion bears a contribution leading to the group velocity breaking the spatial symmetry~\cite{Nembach2015,Gladii2016,Mook2016,Moon2013,Cortes-Ortuno2013,Schlitz2021,Yu2021,Wang2020}. A prominent example of this are the chiral modes that live on the sample edges~\cite{Mook2014,Mohseni2019,Kim2016a,Owerre2016,Thingstad2019}. Such modes have previously been exploited in directional thermal transport~\cite{An2013}, for example. Their role in influencing the magnon chemical potential can thus also be anticipated. Here, we restrict our discussion to the thin film situation which is unable to distinguish between the two surfaces of the magnet, keeping in mind that transport dominated by one surface yields results similar to our analysis below.

We account for the spatial symmetry breaking in the reciprocal space by including a small antisymmetric contribution to the magnon dispersion $\omega(\pmb{k}) = \omega_s(\pmb{k}) + \omega_a(\pmb{k})$, such that $\omega_s(- \pmb{k}) = \omega_s(\pmb{k})$ and $\omega_a(- \pmb{k}) = - \omega_a(\pmb{k})$. Such contributions have been theoretically predicted and experimentally observed in chiral magnets and hybrids~\cite{Nembach2015,Cortes-Ortuno2013,Moon2013,Wang2020,Garst2017,Tokura2018}. The emergence of such a contribution due to the interfacial Dzyaloshinskii-Moriya interaction has been found to cause drift and the consequent control of magnon propagation length in recent nonlocal experiments~\cite{Schlitz2021}. 

In our theoretical analysis here, we account for the antisymmetric contribution perturbatively i.e., we retain terms up to the first order in $\omega_a(\pmb{k})$ and assume that it does not alter the symmetric Brillouin zone of the material. The latter assumption is justifiable when $\omega_a(\pmb{k}) = \pmb{v}_{0} \cdot \pmb{k}$ is caused by a spatial symmetry breaking via an interface perpendicular to $\pmb{v}_0$~\cite{Nembach2015,Kim2016,Moon2013,Wang2020,Schlitz2021}. This is because an alteration of the spatial symmetry along a specific direction should not affect the Brillouin zone properties along orthogonal directions. However, this assumption and the following analysis need not be valid in chiral materials~\cite{Garst2017,Tokura2018} with bulk inversion symmetry breaking giving rise to $\omega_a(\pmb{k})$. An analysis of such materials will be carried out elsewhere.

The magnon current can be evaluated using~(\ref{eq:jm}) with the distribution function in the relaxation time approximation given by:
\begin{align}
f(\pmb{r},\pmb{k}) & = n_{B}(\hbar \omega_s + \hbar \omega_a - \mu) + \delta f(\pmb{r},\pmb{k}),
\end{align}
where
\begin{align}
n_{B}(\hbar \omega_s + \hbar \omega_a - \mu) & \approx  n_{B}(\hbar \omega_s) - \left. \frac{\partial n_{B}}{\partial \epsilon} \right|_{\hbar \omega_s} \mu + \hbar \omega_a \left( \left. \frac{\partial n_{B}}{\partial \epsilon} \right|_{\hbar \omega_s} - \left. \frac{\partial^2 n_{B}}{\partial \epsilon^2} \right|_{\hbar \omega_s} \mu  \right), \\
\delta f(\pmb{r},\pmb{k}) & \approx  - \tau \pmb{v}_g \cdot \pmb{\nabla} n_{B}(\hbar \omega_s + \hbar \omega_a - \mu) = - \tau \left. \frac{\partial n_{B}}{\partial \epsilon} \right|_{\hbar \omega - \mu} (\pmb{v}_s + \pmb{v}_{a})\cdot \pmb{\nabla} \mu,
\end{align}
where $\pmb{v}_{s,a} \equiv \pmb{\nabla}_{\pmb{k}} \omega_{s,a}(\pmb{k})$. Employing the two equations above in~(\ref{eq:jm}) and retaining the lowest order terms, we obtain the desired general expression for the magnon current given in~(\ref{eq:jgen}) with:
\begin{align}
\pmb{j}_0 & = \int_{\mathrm{BZ}} \frac{d^3k}{\left( 2 \pi \right)^3} \left[ n_{B}(\hbar \omega_s) \pmb{v}_a + \left. \frac{\partial n_{B}}{\partial \epsilon} \right|_{\hbar \omega_s} \hbar \omega_a \pmb{v}_s \right], \\
\sigma_m & = - \tau \int_{\mathrm{BZ}} \frac{d^3k}{\left( 2 \pi \right)^3}  \left. \frac{\partial n_{B}}{\partial \epsilon} \right|_{\hbar \omega_s} v_{sz}^2, \\
\chi \pmb{v}_d & = - \int_{\mathrm{BZ}} \frac{d^3k}{\left( 2 \pi \right)^3} \left[ \left. \frac{\partial n_{B}}{\partial \epsilon} \right|_{\hbar \omega_s} \pmb{v}_a + \left. \frac{\partial^2 n_{B}}{\partial^2 \epsilon} \right|_{\hbar \omega_s} \hbar \omega_a \pmb{v}_s  \right].
\end{align}
Considering $\omega_s(\pmb{k}) = \omega_0 + \beta k^2$ and $\omega_a(\pmb{k}) = \pmb{v}_0 \cdot \pmb{k}$, the expressions above are evaluated as:
\begin{align}
\pmb{j}_0 & = \pmb{v}_0 \left( n_0 - \frac{\chi \hbar D_m}{2 \beta \tau} \right), \\
\sigma_m & = \chi D_m,  \\
\pmb{v}_d & =  \pmb{v}_0 \left( 1 - \Gamma \right), \label{eq:vdv0}
\end{align}
where $\chi$, $D_m$, and $\Gamma$ have been defined previously in Eqs.~(\ref{eq:chi}) - (\ref{eq:Gamma}). Further, $n_0$ is the magnon density in equilibrium given by:
\begin{align}\label{eq:magdenasymm}
n_0 & = \int_{\mathrm{BZ}} \frac{d^3k}{\left( 2 \pi \right)^3} n_{B}(\hbar \omega_s) .
\end{align}
Equation (\ref{eq:vdv0}) is the desired expression for the drift velocity and constitutes the main result of this section. The contribution $\Gamma$ has been disregarded in the theoretical analysis of Ref.~\cite{Schlitz2021} and is significant.

One might consider an alternative approach. The $\omega_a(\pmb{k}) = \pmb{v}_0 \cdot \pmb{k}$ contribution considered in evaluating~(\ref{eq:vdv0}) effectively shifts the bottom of the parabolic dispersion away from $\pmb{k} = \pmb{0}$. Thus, one could redefine a new variable $\pmb{\tilde{k}}$ centered around the new dispersion minimum, thereby reducing the problem to the one without the linear-in-k contribution to the dispersion. This is however not allowed because the integration is over the first Brillouin zone [(\ref{eq:jm})]. Any change of variable necessitates modifying the integration domain in a consistent manner. Hence, it is convenient to not make a variable substitution and follow our method of separating even and odd contributions. Further, in contrast with charge currents, global nondissipative spin currents in equilibrium are allowed since spin-orbit coupling or similar effects render spin a non-conserved quantity (e.g., see Ref.~\cite{Chen2014}).

\begin{figure}[tb]
	\begin{center}
		\includegraphics[width=85mm]{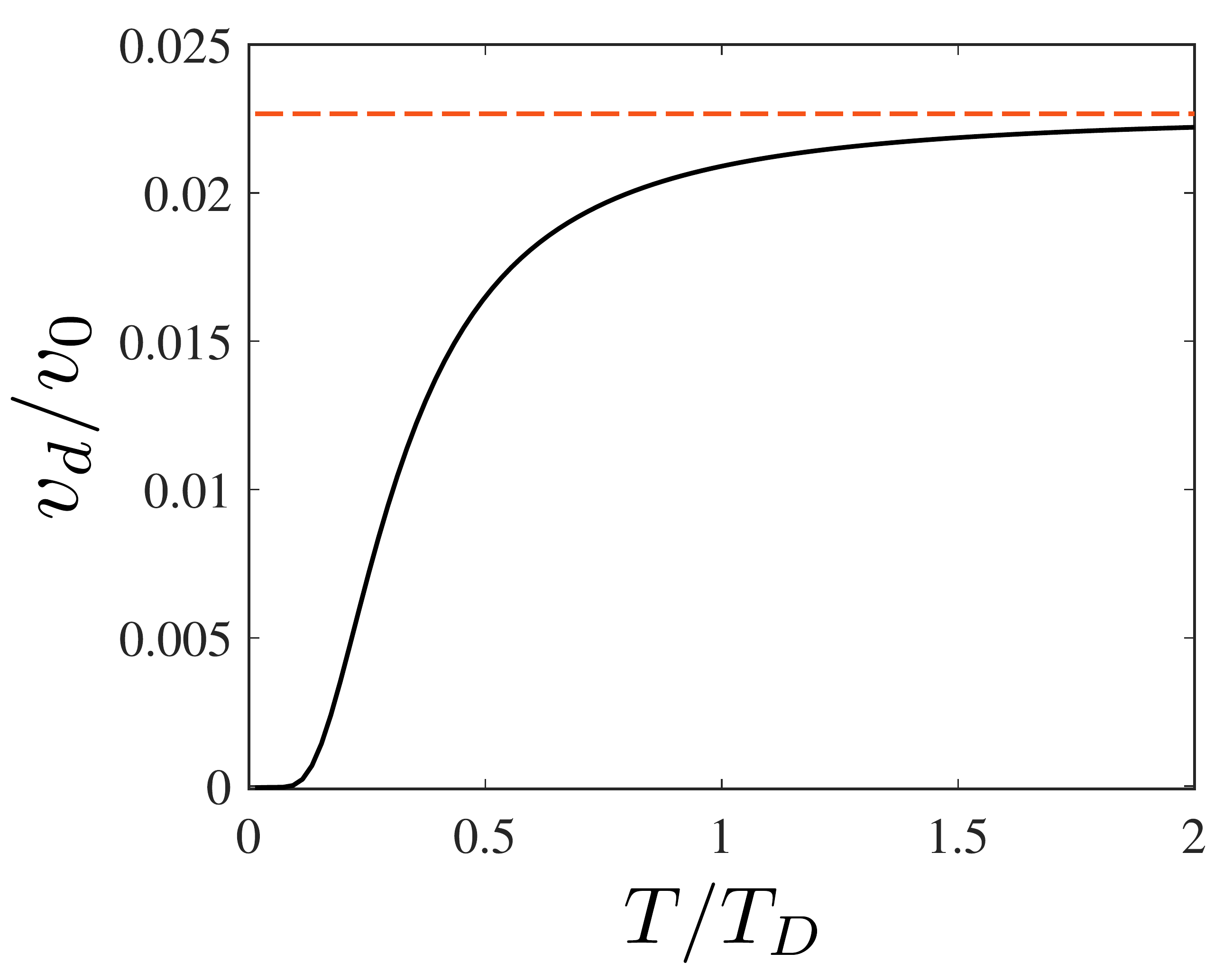}
		\caption{Drift velocity vs.~temperature resulting from an asymmetric contribution $\pmb{v}_0 \cdot \pmb{k}$ to the magnon dispersion [(\ref{eq:vdv0})]. Magnon Debye temperature $T_D = 400$ K and $\hbar \omega_0 / k_B = 1$ K have been assumed. The dashed line indicates the asymptotic value provided in~(\ref{eq:vdv0app}).}
		\label{fig:drift_v0}
	\end{center}
\end{figure}

The ensuing drift velocity [(\ref{eq:vdv0})] has been plotted against temperature in Figure~\ref{fig:drift_v0}, where $\Gamma$ [(\ref{eq:Gamma})] has been evaluated assuming a spherical Brillouin zone with radius $k_{\mathrm{BZ}}$ defined in terms of the magnon Debye temperature $T_D$ assuming a single branch in the dispersion: $\hbar \beta k_{\mathrm{BZ}}^2 \equiv k_B T_{D}$. The low temperature limit of $\Gamma \to 1$, which corresponds to a vanishing drift velocity (Figure~\ref{fig:drift_v0}), has been corroborated via an analytic calculation. The high temperature limit of $\Gamma$ is evaluated as:
\begin{align}\label{eq:asymp}
\Gamma (T \to \infty) & \approx 1 - \frac{2}{3 t \arctan t - 3}, 
\end{align}
where $t \equiv \sqrt{k_B T_D/ \hbar \omega_0} \gg 1$. As shown by the numerically generated plot of Figure~\ref{fig:drift_v0}, this limit is nearly attained for temperatures comparable to $T_D$ and the drift velocity observed at room temperature is close to this value:
\begin{align}
\pmb{v}_d & \approx   \frac{2}{3 t \arctan t - 3} ~ \pmb{v}_0. \label{eq:vdv0app}
\end{align}
Thus, for yttrium iron garnet with the lowest dispersion branch $T_D$ of about 400 K~\cite{Cherepanov1993,Cornelissen2016,Barker2016}, the drift velocity at room temperature is expected to be two orders of magnitude smaller than $v_0$. With $v_0 \sim 10$ m/s~\cite{Wang2020,Schlitz2021}, $v_d \sim 0.1$ m/s is still large compared to what we estimate in the previous section from a reasonable magnetic field gradient. Thus, for the mechanism under consideration, high temperatures are favorable for drift.

\section{Thermal drive}\label{sec:thermal}

The mechanisms for spatial symmetry breaking, and consequently drift, considered above may be classified as ``coherent'' since they stem from altering a single-particle property - the energy. Now, we consider an incoherent mechanism in the form of temperature gradient. A spatial symmetry breaking by a thermodynamic property, such as temperature, describing the ensemble could also lead to drift. Our consideration of thermal gradient for effecting drift is encouraged by the success of thermally driven spin currents in a wide range of magnetic materials and hybrids~\cite{Uchida2010,Bauer2012,Xiao2010,Adachi2011,Uchida2008,Jaworski2010,Adachi2013,Wu2016,Seki2015,Hirobe2017,Rezende2016,Ohnuma2013,Geprags2016,Rezende2014}. However, our considered influence of thermally generated magnon drift on electrically injected and detected magnon transport is qualitatively different from the spin Seebeck effect. For example, in our considerations, no voltage is detected by the normal metal electrode if nonequilibrium magnons are not injected first [e.g., see~(\ref{eq:muevol})], irrespective of the thermal gradient. Further, with the aim of examining drift velocity in the presence of more than one mechanism causing it, we consider the same magnon dispersion $\omega = \omega_s + \omega_a = \omega_0 + \beta k^2 + \pmb{v}_0 \cdot \pmb{k}$, as we did in the preceding section.

The magnon current can be evaluated via~(\ref{eq:jm}) employing the usual relaxation time approximation, as described in \ref{sec:formulate} and used in the previous sections. Furthermore, we continue to assume the magnon and phonon temperatures to be the same, due to the magnon-phonon thermalization time being small ($\sim 1$ ps)~\cite{Cornelissen2016}. In the present evaluation, the equilibrium distribution $f_0(\pmb{r},\pmb{k})$ is assumed to be the Bose distribution function with a spatially homogeneous temperature. In evaluating the magnon current driven by a constant temperature gradient, we assume the deviation of temperature from its assumed equilibrium spatially homogeneous value to be small, and substitute this deviation to zero in our final result. After lengthy calculations and retaining only the lowest order terms, we obtain magnon current density in the form of~(\ref{eq:jgen}) with:
\begin{align}
\pmb{j}_0 & = \pmb{v}_0 \left( n_0 - \frac{\chi \hbar D_m}{2 \beta \tau} \right) - \alpha \pmb{\nabla} T, \label{eq:j0therm} \\
\alpha & \equiv   - \frac{4 \beta^2 \hbar \tau}{T} \int_{\mathrm{BZ}} \frac{d^3k}{\left( 2 \pi \right)^3}  \left. \frac{\partial n_{B}}{\partial \epsilon} \right|_{\hbar \omega_s} k_z^2 \omega_s, \label{eq:alpha} \\
\pmb{v}_d & = \pmb{v}_0 \left( 1 - \Gamma \right) - \Lambda \pmb{\nabla} T, \label{eq:vdtemp} \\
\Lambda & \equiv \frac{4 \beta^2 \hbar \tau}{T \chi}  \int_{\mathrm{BZ}} \frac{d^3k}{\left( 2 \pi \right)^3}  \left. \frac{\partial^2 n_{B}}{\partial \epsilon^2} \right|_{\hbar \omega_s} k_z^2 \omega_s . \label{eq:lambda}
\end{align}
Employing the current expression thus obtained in the magnon continuity equation (\ref{eq:magcont}), we obtain the chemical potential dynamics in the form of~(\ref{eq:chemcont}) with the drift velocity given by~(\ref{eq:vdtemp}). 

In obtaining the results above [Eqs.~(\ref{eq:j0therm})-(\ref{eq:lambda})], we limit ourselves to linear response considering $\pmb{\nabla}T$ and $\pmb{\nabla} \mu$ as the linear drives. The drift term $\propto \mu \pmb{\nabla}T$ is captured adequately in this approximation. In considering nonlocal magnon transport, we are only interested in the transport of magnons that have been injected externally. This means that in our magnon continuity equation, we should only have terms which vanish in the limit $\mu \to 0$. Our linear response results above naturally yield only such terms. If one goes beyond linear response, terms which do not vanish in the limit $\mu \to 0$ and act as source or sink of magnons can be expected. However, in the ongoing analysis, these are neither needed nor can they be faithfully evaluated.

\begin{figure}[tb]
	\begin{center}
		\includegraphics[width=85mm]{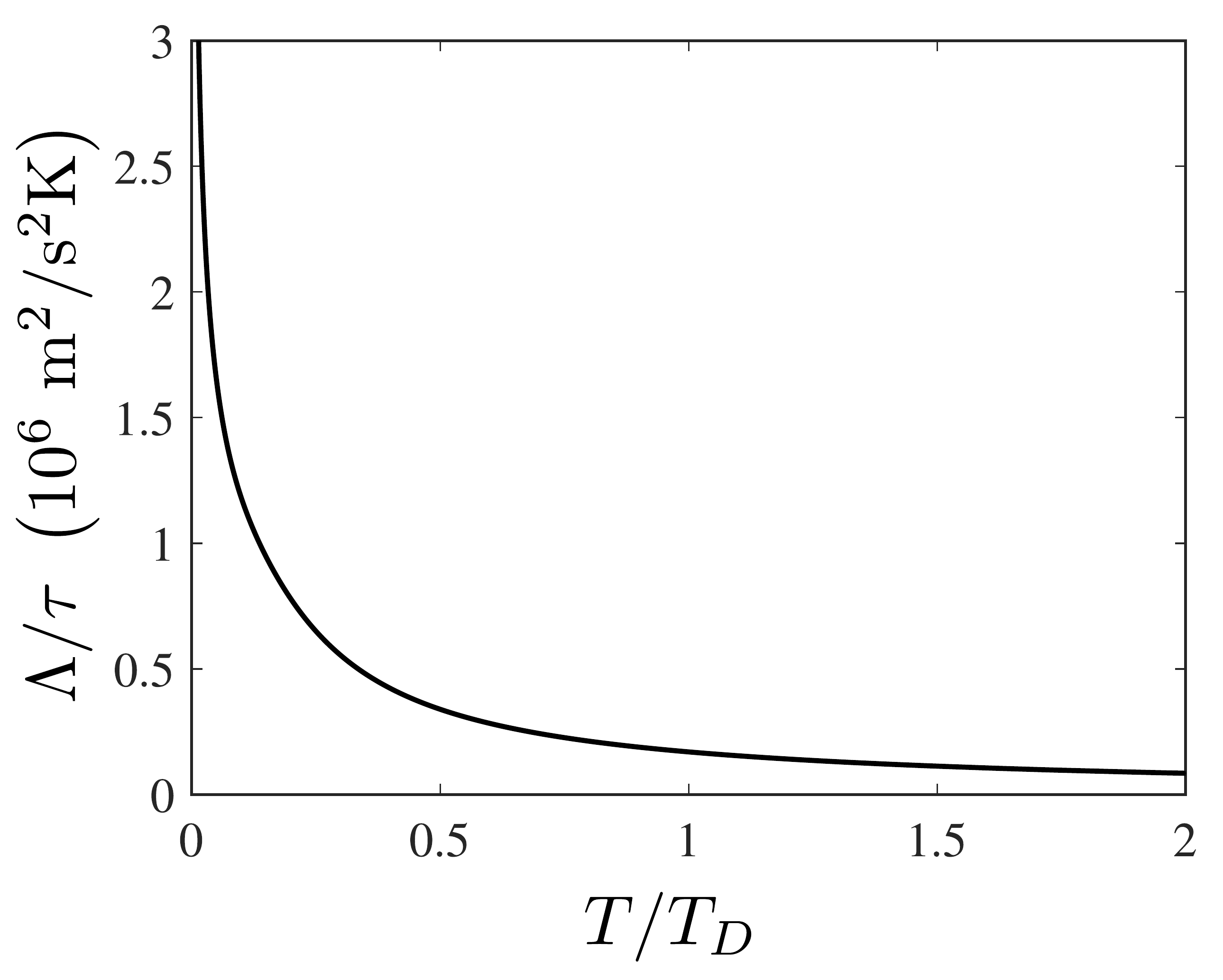}
		\caption{Temperature dependence of the proportionality factor $\Lambda$ that relates drift velocity to the thermal gradient [Eqs.(\ref{eq:vdtemp}) and (\ref{eq:lambda})]. $T_D = 400$ K and $\hbar \omega_0 / k_B = 1$ K have been assumed.}
		\label{fig:Lambda}
	\end{center}
\end{figure}

Equations (\ref{eq:vdtemp}) and (\ref{eq:lambda}) are the main result of the present section and show that the effects of multiple drift mechanisms simply add up in the drift velocity expression. Our explicit demonstration of this here could be anticipated on the basis of linear response principles. The parameter $\Lambda$ [(\ref{eq:lambda})] determines the drift velocity caused by the applied thermal gradient and has been plotted in Figure~\ref{fig:Lambda}. Since $\Lambda$ scales as $1/T$, it decreases with temperature making low temperatures favorable for thermally-induced magnon drift. Furthermore, $\tau$ in the expression for $\Lambda$ [(\ref{eq:lambda})] is not determined by magnon-phonon scattering, since we assume the magnon and phonon temperatures to be identical. The time scale $\tau$ would have been related to magnon-phonon scattering if we would consider a thermal gradient in magnon temperature alone while the phonons would remain in equilibrium. Since we consider a common temperature gradient for both subsystems, $\tau$ is a relatively long time scale governed by scattering with impurities, among others. The value $\tau \sim 1$ ns has been employed previously~\cite{Liao2014} in obtaining results consistent with experiments on yttrium iron garnet, and we adapt this value here. Thus, employing $\tau = 1$ ns, $\hbar \beta = 8\cdot 10^{-40}$ J$\mathrm{m^2}$~\cite{Cornelissen2016}, $T_D = 400$ K~\cite{Cornelissen2016,Cherepanov1993,Barker2016}, $\hbar \omega_0/k_B = 1$ K~\cite{Cornelissen2016}, we obtain $\pmb{v}_d =  (2 \times 10^{-4} ~ \mathrm{m}^2/\mathrm{K} \mathrm{s}) ~ \pmb{\nabla}T$ at room temperature. Thus, drift velocities comparable to those observed in recent experiments~\cite{Schlitz2021} are expected for thermal gradients $\sim 1$ K/mm. In contrast with magnetic field, large temperature gradients are regularly and conveniently generated in labs due to the localized nature of heat generation in solid state systems~\cite{Jaworski2010,Uchida2010,Uchida2008}. Equipped with the above analysis, we expect temperature gradients to offer a convenient and effective handle of magnonic drift velocity and thus, the magnon propagation length.


\section{Conclusion}\label{sec:conclusion}

Inspired by the corresponding role in electronic spin transport~\cite{Fabian2007} and recent experiments on magnons~\cite{Schlitz2021}, we have theoretically examined the role of drift currents in controlling the magnonic spin propagation length, as observed in nonlocal experiments. Employing a phenomenological description based on the random walk model~\cite{Fabian2007}, we established that a magnon drift current requires spatial symmetry breaking in the magnonic description. We further found that it modulates the magnon propagation length, which can be a direct experimental signature, as established in recent experiments~\cite{Schlitz2021}. Employing the Boltzmann approach, we evaluated the magnon drift velocity caused by three representative symmetry breaking mechanisms - magnetic field gradient, magnon chirality, and thermal gradient. Using experimentally measured parameters (which vary considerably among different reports), we estimate these drives to potentially result in an increasing magnitude of the drift velocity, with thermal drive potentially offering the largest values. 

Our analysis here assumes a quasi-continuous magnon dispersion and evaluates the transport coefficients, especially the drift velocity, in terms of the Brillouin zone sum over all magnon modes. While providing useful general relations, the present endeavor will benefit from a more microscopic and numerical analysis in the future, where a spin Hamiltonian is instead the starting point. This will be especially important for examining the case of asymmetric contribution to the magnon dispersion with regards to its spatial dependencies and the Brillouin zone rigidity. On the experimental side, our analysis anticipates and paves the way for employing thermal drive, and the consequent drift currents, in controlling the magnon propagation length in nonlocal magnonic spin transport experiments.


\section*{Acknowledgments}
We thank Sebastian Goennenwein, Bart van Wees, and Gerrit Bauer for valuable discussions. We acknowledge financial support from the Spanish Ministry for Science and Innovation -- AEI Grant CEX2018-000805-M (through the ``Maria de Maeztu'' Programme for Units of Excellence in R\&D) and grant PID2020-114880GB-I00. Sebastián de-la-Peña is grateful to Instituto Universitario de Ciencia de Materiales Nicolás Cabrera for one of its Research Awards for Physics Students 2021.


\appendix

\section{Formulation of magnon continuity and constitutive equations}\label{sec:formulate}

In this section, we derive our general methodology followed in investigating the magnon chemical potential dynamics. Making use of the fact that two very different time scales govern the magnon number decay and their transport~\cite{Cornelissen2016}, we formulate magnon continuity equation at long time scales. On the other hand, constitutive equations for the system in the form of current-drive relations are derived by solving the Boltzmann equation at short time scales~\cite{Chen2005,Cornelissen2016,Kittel2004}. 

For simplicity, we consider only two time scales that suffice for the problem at hand. $\tau$ is a small time governed by the magnon number conserving scattering events mediated by exchange interaction~\cite{Akhiezer1968,Cornelissen2016}, while $\tau_m$ is the magnon density decay time due to the scattering processes mediated by relativistic effects, such as spin-orbit coupling and dipolar interactions~\cite{Akhiezer1968,Cornelissen2016}. The Boltzmann equation is written as~\cite{Kittel2004}:
\begin{align}
\frac{\partial f}{\partial t} + \pmb{v}_g \cdot \pmb{\nabla} f & = \left. \frac{\partial f}{\partial t} \right|_{\mathrm{coll}},
\end{align}
where $f \equiv f(\pmb{r},\pmb{k},t)$ is the distribution function, $\pmb{v}_g \equiv \pmb{\nabla}_{\pmb{k}} \omega$ is the group velocity, and the right hand side represents the collision integral. Within our two-times model, we may express the collision integral as the sum over magnon-number conserving term and magnon-decay term~\cite{Cornelissen2016}:
\begin{align}
\frac{\partial f}{\partial t} + \pmb{v}_g \cdot \pmb{\nabla} f & = \left. \frac{\partial f}{\partial t} \right|_{\mathrm{cons}} +  \left. \frac{\partial f}{\partial t} \right|_{\mathrm{decay}}. \label{eq:boltz1}
\end{align}
To formulate the magnon continuity, we integrate the equation above over the whole Brillouin zone (BZ):
\begin{align}
\int_{\mathrm{BZ}} \frac{d^3k}{\left( 2 \pi \right)^3} \frac{\partial f}{\partial t} + \int_{\mathrm{BZ}} \frac{d^3k}{\left( 2 \pi \right)^3} \pmb{v}_g \cdot \pmb{\nabla} f & = \left\{ \int_{\mathrm{BZ}} \frac{d^3k}{\left( 2 \pi \right)^3} \left. \frac{\partial f}{\partial t} \right|_{\mathrm{cons}} \right\} +  \int_{\mathrm{BZ}} \frac{d^3k}{\left( 2 \pi \right)^3} \left. \frac{\partial f}{\partial t} \right|_{\mathrm{decay}}, \label{eq:magcontint} \\
\frac{\partial n(\pmb{r})}{\partial t} + \pmb{\nabla} \cdot \pmb{j}_m & = \int_{\mathrm{BZ}} \frac{d^3k}{\left( 2 \pi \right)^3} \left. \frac{\partial f}{\partial t} \right|_{\mathrm{decay}}, \label{eq:magcontapp}
\end{align}
where the magnon density $n(\pmb{r})$ and current density $\pmb{j}_m$ (assuming that $\pmb{v}_g$ does not depend on the spatial coordinate $\pmb{r}$) are given by:
\begin{align}
n(\pmb{r}) & = \int_{\mathrm{BZ}} \frac{d^3k}{\left( 2 \pi \right)^3} f , \label{eq:n} \\
\pmb{j}_m & = \int_{\mathrm{BZ}} \frac{d^3k}{\left( 2 \pi \right)^3} \pmb{v}_g f. \label{eq:curr}
\end{align}
In~(\ref{eq:magcontint}), the term in curly brackets vanishes due to its magnon conserving nature. This part of the collision integral redistributes magnons between different $\pmb{k}$ values but does not change their total number. Equation (\ref{eq:magcontapp}) is our general formulation of magnon continuity, which is nearly assumption free. Making the relaxation time approximation for magnon decay via~\cite{Cornelissen2016}:
\begin{align}
\left. \frac{\partial f}{\partial t} \right|_{\mathrm{decay}} & = - \frac{f - n_B(\hbar \omega)}{\tau_m},
\end{align}
leads us to the continuity equation (\ref{eq:magcont}) valid at long time scales and employed in the main text.

To complete the analysis under a given drive, we need to evaluate the current density using~(\ref{eq:curr}) and the Boltzmann equation (\ref{eq:boltz1}) on short time scales:
\begin{align}
\frac{\partial f}{\partial t} + \pmb{v}_g \cdot \pmb{\nabla} f & = \left. \frac{\partial f}{\partial t} \right|_{\mathrm{cons}}, \label{eq:boltzshort}
\end{align}
where the small magnon decay term is disregarded. We further employ the relaxation time approximation:
\begin{align}
\left. \frac{\partial f}{\partial t} \right|_{\mathrm{cons}} & = -  \frac{f - n_B(\hbar \omega - \mu)}{\tau},
\end{align}
where the assumed $\mu$ needs to be determined by solving the more general spin-continuity equation (\ref{eq:magcont}). The relaxation time approximation is not adequate for capturing certain features of the collision integral~\cite{Chen2005,Liao2014,Sanders1977}. Thus, we need to validate our results obtained within the relaxation time approximation against the expected fundamental properties, e.g., spin conservation on small time scales. In particular, $\tau$ should be considered a phenomenological time scale and needs to be determined for a given situation based on physical arguments and comparison with experiments~\cite{Chen2005,Liao2014}. In steady state, the perturbative solution for $f$ using the above equations is obtained as:
\begin{align}
f & = n_{B}(\hbar \omega - \mu) - \tau \pmb{v}_g \cdot \pmb{\nabla} n_{B}(\hbar \omega - \mu),
\end{align} 
where the right hand side takes different forms depending on the drive considered. Substituting the above perturbative solution to $f$ in~(\ref{eq:curr}) yields the desired current-drive constitutive relations for the system.

\section{Nonlocal magnetoresistance and magnon chemical potential}\label{sec:Rnl}

We now relate the experimentally measured nonlocal magnetoresistance with the magnon chemical potential and its spatial evolution within a simplified model. We assume the injector and detector electrodes to be weakly coupled to the magnet. This assumption is made to simplify our theoretical analysis and to ascertain the essential physics quantitatively within a rigorous model. The experiments like to have a measurable magnetoresistance which benefits from a stronger coupling between the magnet and heavy metal electrodes. This entails a more complicated inclusion of the boundary conditions~\cite{Cornelissen2016}, which can eclipse the key phenomena at play. Hence, our simplified model clarifies the main qualitative physics and should be understood in that manner. The terms that result purely from our boundary conditions should not be relied on for comparison with experiments. We note that the control of magnon propagation length [(\ref{eq:lp})] does not depend on boundary conditions. 

The key goals and messages of the following analysis are (i) to justify the boundary condition~(\ref{eq:bc1}), (ii) to show that the nonlocal magnetoresistance is proportional to the magnon chemical potential at the detector location, and (iii) to determine the magnitude and scaling of the nonlocal magnetoresistance. We follow an analysis that has been detailed under more general conditions elsewhere~\cite{Cornelissen2016,Guckelhornarxiv}. Here, for ease of notation, we assume the injector and detector electrodes to be identical in terms of materials and physical dimensions. 

Under the assumption of small spin conductance of the heavy metal-magnetic insulator interface, the spin accumulation (in units of energy) generated in the injector electrode by a charge current density $j_{ci}$ driven through it is~\cite{Nakayama2013,Cornelissen2016}:
\begin{align}\label{eq:kappa}
\mu_{s,\mathrm{inj}} & = 2 e \theta l_{s} \rho \tanh \left( \frac{t}{2 l_{s}} \right) j_{ci} \equiv \kappa j_{ci},
\end{align}
where $\theta$ is the spin Hall angle, $e$ is the electronic charge magnitude, $\rho$ is the resistivity, $l_s$ is the spin diffusion length, and $t$ is the electrode thickness. This injects a magnonic spin current into the magnet~\cite{Bender2012}:
\begin{align}\label{eq:Iinj}
I_{s,\mathrm{inj}} & = g_{\mathrm{int}} w L \left( \mu_{s,\mathrm{inj}} - \mu \right) \approx g_{\mathrm{int}} w L  \mu_{s,\mathrm{inj}},
\end{align}
where $g_{\mathrm{int}}$ is the interfacial spin conductance per unit area, $w$ is the injector width assumed to be much smaller than the spin propagation length $\lambda_m$, and $L$ is the injector length assumed uniform throughout the whole device including the magnetic insulator layer. The approximation above is valid under the limit $g_{\mathrm{int}} w L \to 0$ such that the entire spin potential drops across the highly resistive interface~\cite{Cornelissen2016}. Considering that half of the magnonic spin current injected propagates towards the positive direction and contributes directly to the magnon chemical potential gradient, we have derived the boundary condition~(\ref{eq:bc1}):
\begin{align}
- \left. \sigma_m \frac{\partial \mu}{\partial x} \right|_{x=0} & = j_{\mathrm{in}} = \frac{g_{\mathrm{int}} w   \kappa}{2 t_m} j_{ci},
\end{align}
where $t_m$ is the magnet thickness and the injector is assumed infinitesimal in size and located at $x = 0$. This accomplishes our first goal mentioned above regarding the boundary condition. 

Employing analysis similar to~(\ref{eq:Iinj}), we evaluate the spin current absorbed by the detector electrode as:
\begin{align}
I_{s,\mathrm{det}} & = g_{\mathrm{int}} w L \left( \mu(x_d) - \mu_{s,\mathrm{det}} \right) \approx g_{\mathrm{int}} w L \mu(x_d),
\end{align}
where the detector is assumed to be located at $x = x_d$. Thus, we see that the spin current absorbed and the detected signal is governed by the magnon chemical potential at the detector location: $\mu(x_d)$, thereby accomplishing our second goal mentioned above. 

Finally, the voltage generated in the detector electrode~\cite{Mosendz2010} due to the absorbed spin current can be evaluated in a manner which is the reverse process of deriving~(\ref{eq:kappa}). We do not detail it here, but directly write the result:
\begin{align}
V_{\mathrm{det}} & = \frac{\kappa g_{\mathrm{int}} L}{\hbar t} \mu(x_d).
\end{align}
Employing the equations above, we obtain the desired expression for nonlocal magnetoresistance:
\begin{align}\label{eq:Rnl}
R_{\mathrm{nl}} & = \frac{V_{\mathrm{det}}}{I_{ci}} = \frac{2 L}{\hbar} \ \frac{\kappa^2}{4 t^2} \ \frac{l_{p}}{D_m t_m} \ \left( \frac{g^2_{\mathrm{int}}}{\chi} \right) \ \tilde{\mu}(x_d),
\end{align}
where $I_{ci} = j_{ci} w t$ is the charge current driven through the injector, we employed the relation $\sigma_m = D_m \chi$, and $l_p$ is the magnon propagation length. Here, employing~(\ref{eq:muxd}), $\tilde{\mu}(x_d)$ is the magnon chemical potential normalized to its value at the injector $x_d = 0$ and captures its spatial dependence.

Equation (\ref{eq:Rnl}) accomplishes our third and final goal of establishing the dependence of the experimentally recorded nonlocal magnetoresistance. The dominant temperature dependence of $R_{\mathrm{nl}}$ comes from the factor enclosed in brackets [(\ref{eq:Rnl})]. It increases with magnon population and temperature since both $g_{\mathrm{int}}$ and $\chi$ increase with the magnon population. Hence, the experimentally observed signal diminishes at low temperatures. Further, $g_{\mathrm{int}}$ is unaffected by the asymmetric contribution to the dispersion considered in Section~\ref{sec:reciprocal} up to the first order in $\omega_a$, similar to the magnon density [(\ref{eq:magdenasymm})].


\bibliography{MagDrift}

\end{document}